# Micro-X Sounding Rocket: Transitioning from First Flight to a Dark Matter Configuration

J. S. Adams[1] · A. J. Anderson[1] · R. Baker[1] · S. R. Bandler[1] · N. Bastidon[1] ·
D. Castro[1] · M. E. Danowski[1] · W. B. Doriese[1] · M. E. Eckart[1] ·
E. Figueroa-Feliciano[1] · D. C. Goldfinger[1] · S. N. T. Heine[1] · G. C. Hilton[1] ·
A. J. F. Hubbard[1] · R. L. Kelley[1] · C. A. Kilbourne[1] ·
R. E. Manzagol-Harwood[1] · D. McCammon[1] · T. Okajima[1] · F. S. Porter[1] ·
C. D. Reintsema[1] · P. Serlemitsos[1] · S. J. Smith[1] · P. Wikus[1]

## Abstract

The Micro-X sounding rocket flew for the first time on July 22, 2018, becoming the first program to fly Transition-Edge Sensors and multiplexing SQUID readout electronics in space. While a rocket pointing failure led to no time on-target, the success of the flight systems was demonstrated. The successful flight operation of the instrument puts the program in a position to modify the payload for indirect galactic dark matter searches. The payload modifications are motivated by the science requirements of this observation. Micro-X can achieve world-leading sensitivity in the keV regime with a single flight. Dark matter sensitivity projections have been updated to include recent observations and the expected sensitivity of Micro-X to these observed fluxes. If a signal is seen (as seen in the X-ray satellites), Micro-X can differentiate an atomic line from a dark matter signature.

## 1 The Micro-X Payload

The Micro-X X-ray sounding rocket was launched on July 22, 2018 from the White Sands Missile Range in New Mexico, USA. This was the first operation of Transition-Edge Sensors (TES) and their time-division multiplexing (TDM) readout electronics in space, advancing their technology readiness level (TRL) and opening up sensitivity to new physics [1]. The science goal of the first flight was a high-resolution observation of

✉ A. J. F. Hubbard
  ahubbard@northwestern.edu

1 Department of Physics and Astronomy, Northwestern University, Evanston, IL 60208, USA

the Cassiopeia A Supernova Remnant. The Micro-X observation requires a minimum altitude of 160 km so that X-rays in the bandpass are not attenuated by the atmosphere. Each flight provides a 338 s (5.6 min) exposure above 160 km.

## 1.1 Science Instrument

The instrument can be built up in two configurations: an imaging configuration for targets that require spatial resolution and a large-field of view (FOV) configuration to observe diffuse, all-sky signals that do not require spatial resolution. The imaging configuration includes an X-ray optic and is used for Supernova Remnant (SNR) observations; this was the first flight configuration. The large-FOV configuration, described in Sect. 2.1, is designed for galactic dark matter searches, described in Sect. 2.2.

The heart of the science instrument in both configurations is the 128-pixel TES microcalorimeter array. Each 590 µm × 590 µm pixel has a Au/Bi absorber (3.4 µm Bi, 0.6 µm Au) and a Mo/Au TES. A TDM Superconducting QUantum Interference Device (SQUID) readout is used, with 8 columns of 16 rows each [2]. To track detector response in flight, an onboard radioactive calibration source is flown.

The detectors are flown in a cryostat, shown in Fig. 1, and an Adiabatic Demagetization Refrigerator (ADR) is used for temperature control. The ADR must regulate to 75 mK ± 10 µK within 60 s after powered flight for optimal exposure [3]. A set of thin (100 s Å) optical/infrared Al-polyimide filters block external light while transmitting X-rays.

Details on the detectors and their first flight performance are presented in [1]. With the successful demonstration of the detector, cryogenics, and electronics systems

**Fig. 1** The Micro-X cryostat, which opens to space on the bottom side and interfaces to onboard electronics on the top side (Color figure online)

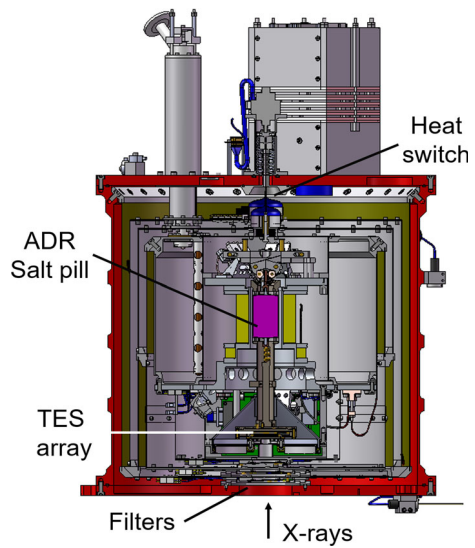

Heat switch

ADR
Salt pill

TES
array

Filters    ↑ X-rays

during the first flight and a reflight scheduled for December 2019, in this paper we focus on our next science target: dark matter.

## 2 Dark Matter Instrument Configuration

### 2.1 Hardware Modifications for a Dark Matter Search

The dark matter instrument goals, shown in Table 1, motivate a new TES array and modifications to the filters and calibration source to optimize for the higher-energy bandpass. The optimization of effective area to resolution for the new array is ongoing. The reported projections assume 128 pixels at $890 \times 890\,\mu m^2$ for an effective area of $1.1\,cm^2$. The proposed detectors are a Mo/Au bilayer TES with a Au/Bi absorber ($3\,\mu m$ Bi, $0.7\,\mu m$ Au). The expected signal rate is $< 10$ Hz across the array ($< 1$ Hz/pixel), which is well within the readout capabilities of the system. The higher-energy bandpass allows the optical/infrared filters to be made thicker than the Athena and XQC filters, which are kept thin to accommodate sub-0.5 keV X-rays. The onboard calibration source will use $^{55}$Fe to fluoresce NaCl (instead of the current KCl) to emit outside the region of interest.

The large FOV requires mechanical modifications to the aft-end apertures and the magnetic shielding. The modifications will widen all apertures between the detector array and space. The detectors are magnetically sensitive, and magnetic shielding for the detectors will be modified for the larger aperture. Current shielding uses a bucking coil between the ADR and the detectors, a Nb shield enclosing the TESs and SQUIDs, and a field coil above the array to reject any remaining field. The imaging configuration of the payload flies a set of magnetic brooms in the optics section to deflect charged particles from the detector array. The impact of incident charged particles for the larger FOV configuration, which will not fly magnetic brooms, is under investigation.

### 2.2 Projected Dark Matter Limits

The successful demonstration of the flight systems in the first flight puts the program in a position to pursue the large-FOV instrument configuration optimized for an indirect galactic dark matter search [4]. Dark matter makes up 26% of the known Universe, but its nature is still unknown [5]. In the Milky Way, a spherical dark matter halo surrounds the plane of luminous matter and produces an all-sky signal as seen from the solar system. Indirect dark matter experiments search for secondary particles that may be produced in dark matter interactions, including X-rays. The X-ray production mechanism depends on the dark matter candidate. If dark matter is a sterile neutrino, it could decay to an X-ray and an active neutrino in a loop-suppressed process. This would produce an X-ray at half the sterile neutrino mass [6]. A keV-scale X-ray would thus result from the decay of a keV-scale sterile neutrino.

**Table 1** Micro-X instrument specifications

| Configuration | Imaging | Dark matter |
|---|---|---|
| Operating temperature | 75 mK | 75 mK |
| Energy resolution (FWHM) | 4.5–10 eV | 3 eV |
| Bandpass | 0.2 4 keV | 0.5 10 keV |
| Effective area | $0.47\,cm^2$ | $1.1\,cm^2$ |
| Pixel size | $590 \times 590\,\mu m^2$ | $890 \times 890\,\mu m^2$ |
| Field of view | 11.8 arcmins | 33° |
| Observation time | 300 s | 300 s |
| Expected counts | 13,000 | 2400 |

Indirect dark matter detection in the X-ray band is particularly compelling because several X-ray satellites (Chandra, XMM-Newton, Suzaku, and NUSTAR) continue to observe an X-ray anomaly at 3.5 keV that may be the secondary product of a dark matter interaction [7,8, and ref. therein]. Observations of the line have been reported from the Galactic Center, galaxies, and galaxy clusters. It does not appear to be an instrumental effect because it is seen in multiple detectors, and the line redshifts with the astronomical target [7,8]. The line may be of atomic origin, but this implies that our predicted fluxes at this energy are off by an order of magnitude [7–9]. Significantly complicating our understanding of this signal are the competing analyses that have not seen the signal, sometimes even using the same data [7,10]. This line is a hotly contested claim that is constrained by systematic and calibration limitations of current instruments. It requires a new, high-resolution detector to be resolved.

A 5-min sounding rocket flight with high-resolution detectors and a large FOV is well-suited to this observation. Galactic dark matter is an all-sky signal, so the incident flux scales with FOV. A 33° half-angle FOV instrument observes a dark matter flux that is 2700 times that of XMM-Newton [4], making up for the short sounding rocket exposure. Microcalorimeters provide the high resolution required to separate the monoenergetic signal from the background continuum and nearby atomic lines, as demonstrated by the XQC and Hitomi observations used to search for this line [4,11]. Modifying Micro-X for a large FOV makes it an excellent instrument for this observation.

The expected flux from galactic dark matter depends on the direction of the observation, the FOV of the instrument, the integrated column density of the dark matter, and the implied decay rate ($\Gamma$) of the dark matter particle. A Navarro–Frenk–White (NFW) dark matter density profile [12] was used to normalize previously observed fluxes taken in different directions to the Micro-X target fields:

$$\rho_{\mathrm{NFW}} = \frac{\rho_0}{\left(\frac{r}{r_{\mathrm{s}}}\right)^{\gamma} \left(1 + \frac{r}{r_{\mathrm{s}}}\right)^{3-\gamma}} \quad , \tag{1}$$

where $r$ is the distance from the Galactic Center, $\gamma = 1.0$, the density scale $\rho_0 = 8.54 \times 10^{-6}\,\mathrm{M_{\odot}/kpc^3}$ (solar mass per kiloparsec$^3$), and the scale radius $r_{\mathrm{s}} = 19.6\,\mathrm{kpc}$ [12]. The distance from the Sun to the Galactic Center is assumed to be 8.21 kpc [12].

The value of $\Gamma$ is an ongoing point of discussion, driven by differences in empirically derived best-fit values. Sensitivity projections change with the expected flux (and therefore, decay rate) from a reported dark matter line, and three different decay rates (see Table 2) are explored. All projections use the NFW profile from [12] to normalize to the expected flux to the Micro-X fields. The first is derived from the reported 3.53 keV flux from the Galactic Center with XMM-Newton [13] and used in previous Micro-X sensitivity projections [4], updated using a newer NFW profile [12]. The second is derived by matching the best-fit value of the 3.505 keV flux from Chandra observations of the COSMOS/CDFS fields ($\ell = 223.6°$, $b = −54.4°$) [14]. The third is the best-fit $\Gamma$ reported from a surface brightness profile study across the Milky Way, divided into 5 regions out to 2100' [15], which predicts a significantly smaller flux at 3.494 keV [16].

The empirically driven flux predictions differ by a factor of 6.6, and this discrepancy is not understood; each profile could be correct. These results are limited by the low signal to noise and systematic uncertainties of the instruments. For example, in the [13] analysis of 1.4 Ms of XMM-Newton data there are $\sim 7500$ signal counts in the claimed 3.54 keV line in each MOS detector. Due to the $\mathcal{O}(100)$ eV resolution of these CCD instruments, there are upwards of 500,000 background counts in that resolution element. With a signal-to-noise ratio of 0.015, the analysis relies on an accurate model of the background and depends strongly on the understanding of the instrument's systematic errors to return a high significance ($5.7\sigma$) result. The discrepancies in predicted flux in Table 2 cannot be resolved with current orbiting instruments; resolving this uncertainty is the main science driver for the proposed Micro-X observations.

Two observation fields are proposed, shown in Fig. 2. The first flight (Micro-X North) is proposed to launch from New Mexico and observe a region near the Galactic Center ($\ell = 31$, $b = 40$) that is relatively quiet in the X-ray band. The X-ray background spectrum for this field is relatively flat at 0.6 counts/flight/2.5 eV [4]. The second flight (Micro-X South) is proposed to launch from the Southern Hemisphere and observe the Galactic Center ($\ell = 0, b = −12$). The background spectrum for this field, although higher due to contributions from low-mass X-ray binaries (LMXBs) in the galactic plane, is equally flat at 3.5 counts/flight/2.5 eV [4]. Both fields avoid the bright X-ray source SCO X1. The projected spectra are shown in Fig. 3.

A dark matter flux consistent with [13] or [14] would yield an observation of $> 5\sigma$ significance from a single flight observing either the Micro-X North or the Micro-X South fields (Fig. 4). A dark matter flux consistent with [15] would yield a $> 2\sigma$ observation from Micro-X North and from Micro-X South, and $> 3\sigma$ from the combined dataset. Even with the most conservative flux estimate, Micro-X will achieve $> 3\sigma$ sensitivity with the combination of the proposed flights. The Micro-X observations are statistics-limited, so adding additional flights increases the sensitivity even further.

**Table 2** Dark matter decay rates predicted by previous observations, and their associated projections for the Micro-X north ($\ell = 31$, $b = 40$) and south ($\ell = 0$, $b = −12$) fields, given the detector parameters in Table 1. The expected background rate is 0.6 (3.5) counts/flight/2.5eV bin for the north (south) field

| Data references | DM decay rate ($\Gamma$) (s$^{-1}$) | Micro-X North [counts/flight] (significance) | Micro-X South [counts/flight] (significance) | Micro-X combined (counts) (significance) |
|---|---|---|---|---|
| [13] | $1.52 \times 10^{-28}$ | $20.3 \pm 4.5 \, (> 5\sigma)$ | $37.6 \pm 6.1 (> 5\sigma)$ | $57.9 \pm 7.6 \, (> 5\sigma)$ |
| [14] | $2.59 \times 10^{-28}$ | $29.9 \pm 5.5 \, (> 5\sigma)$ | $55.3 \pm 7.4 (> 5\sigma)$ | $85.2 \pm 9.2 \, (> 5\sigma)$ |
| [15] | $0.39 \times 10^{-28}$ | $4.0 \pm 2.0 \, (> 2\sigma)$ | $7.4 \pm 2.7 (> 2\sigma)$ | $11.4 \pm 3.4 \, (> 3\sigma)$ |

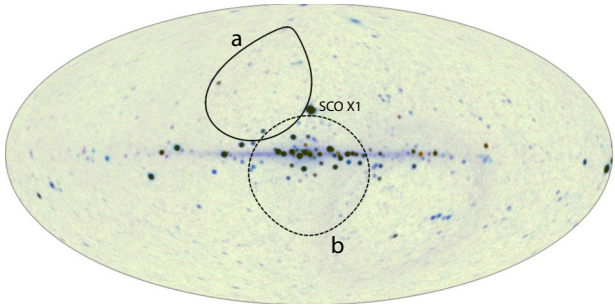

**Fig. 2** The Micro-X North (a) and South (b) fields (Color figure online)

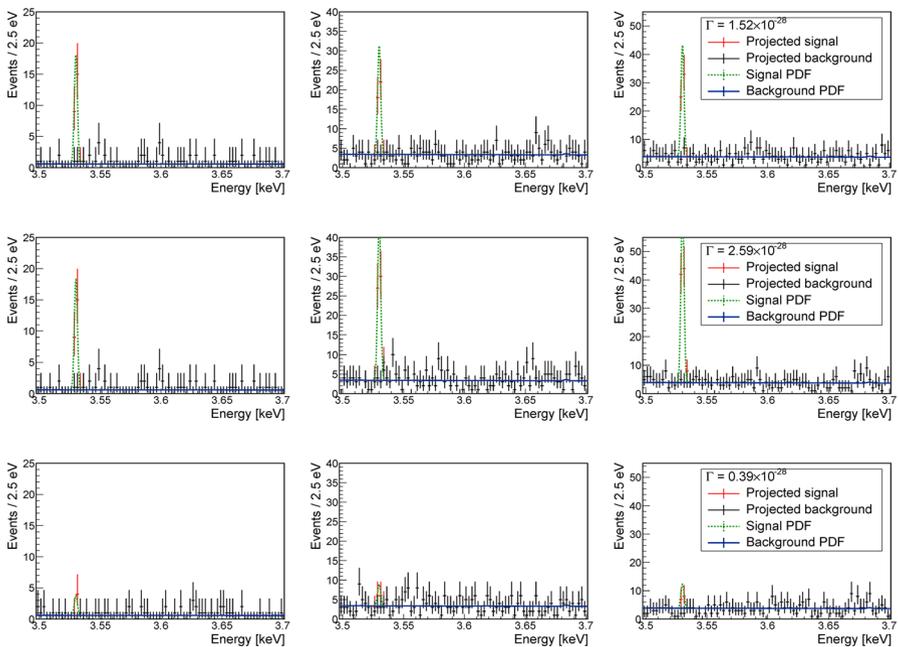

**Fig. 3** Projected spectrum for the Micro-X North (Left), the Micro-X South (Center), and the combined (Right) observations. Projections are shown for the fluxes derived from: [13] (Top), [14] (Middle), and [15] (Bottom). The simulated background data are in black, with the simulated signal in red. The background fit is in blue, with the signal PDF in green. Background in the Micro-X North field is dominated by the cosmic X-ray background (CXB), while for the Micro-X South field it is dominated by LMXBs from the galactic disk (Color figure online)

The dark matter decay rate ($\Gamma$) parameter is independent of the exact dark matter particle candidate. In the case of sterile neutrino decays, $\Gamma$ is given by:

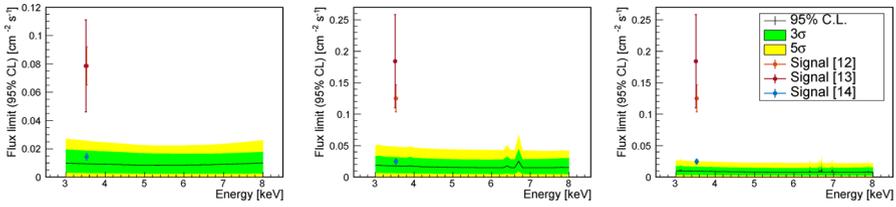

**Fig. 4** Projected dark matter flux limits from: Micro-X North (Left), Micro-X South (Center), and the combined exposure (Right). The 3σ band (corresponding to the distribution of confidence limits for a set of identical experiments) is in green, and the 5σ band is in yellow. The observations from [13], [14], [15] are shown in red, maroon, and blue, respectively (Color figure online)

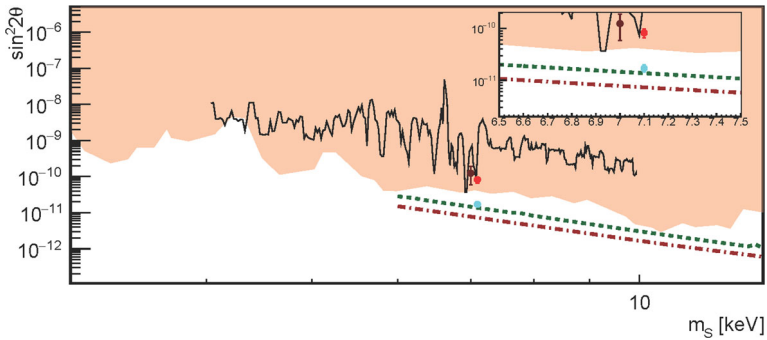

**Fig. 5** Projected exclusion limits for sterile neutrino dark matter from Micro-X North (green dotted) and Micro-X South (red dash-dotted). Previous claims from [13–15] are shown in red, maroon, and blue dots, respectively. Previous exclusion limits from X-ray observations (shaded) and XQC (black line [4]) are shown (Color figure online)

$$\Gamma = \left(1.38 \times 10^{29} s^{-1}\right) \times \left(\frac{\sin^2 2\theta}{10^{-7}}\right) \left(\frac{m_\mathrm{s}}{1\,\mathrm{keV}}\right)^5 \; , \qquad (2)$$

where $m_\mathrm{s}$ is the sterile neutrino mass and $\theta$ is the mixing angle between the active and sterile states [17]. The sterile neutrino parameters derived from the decay rates in Table 2 are shown in Fig. 5.

If a line is detected, it may be from a dark matter interaction or of atomic origin. Micro-X has the ability to discern between these two signals by mapping the Doppler shift of the line across the Galaxy with multiple flights [18,19]. This analysis provides a "smoking gun" signature of dark matter because an atomic background would be co-moving with the Earth in the galactic plane, while a dark matter signal would be coming from the stationary dark matter halo. This velocity spectroscopy requires 0.1% resolution and drives the 3 eV energy resolution specification of the new array.

The Micro-X observation is highly complementary with future XRISM [20] observations. Micro-X is optimized for the all-sky galactic signal, and its short flight precludes it from observing fainter targets that require a longer exposure with XRISM. XRISM will get excellent spectra from extragalactic sources like galaxy clusters and dwarf spheroidals. If XRISM was to observe the Galactic Center, it would take 35 (65) Ms of observation time to accumulate the same signal flux as the Micro-X North (South) field. In a more realistic scenario where multiple observations throughout the lifetime of the mission are co-added (with a distribution of pointing directions through the Milky Way halo), it would take more than 100 Ms of data to accumulate the same signal flux as either Micro-X flight. Thus, a combination of data from both experiments is an excellent way to enhance sensitivity to dark matter signals in the X-ray band.

## 3 Conclusions

With its maiden flight, Micro-X became the first program to fly TES and TDM SQUID readouts in space. The first flight demonstrated the successful engineering and operational performance of the instrument subsystems and allowed the instrument to be flight-tested, despite the rocket pointing failure. The program will re-fly in December 2019 in its original configuration to perform the observation intended for the first flight. It will then be in a position to begin the modifications for an indirect galactic dark matter observation.

To achieve the dark matter science goals, the instrument will be modified for the higher-energy bandpass and wider FOV. A new TES array will be made, and modifications to the apertures, filters, calibration source, and magnetic shielding will be made.

Micro-X is projected to set world-leading limits for keV-scale galactic dark matter signals with a single flight. The expected dark matter flux in the galaxy is under intense study, with multiple observations returning competing flux results. The difference between these fluxes is not understood and requires a high-resolution instrument to resolve. For the traditional profile from [13,14], Micro-X will return a $> 5\sigma$ result in a single flight; even with the more conservative flux prediction from [15], Micro-X will return a $> 3\sigma$ result after the two proposed flights. If the line is observed, Micro-X will be able to identify the line as a dark matter signal rather than an atomic background by pursuing velocity spectroscopy over multiple flights.

**Acknowledgements** We gratefully acknowledge the technical support of Travis Coffroad, Ken Simms, Ernie Buchanan, Tomomi Watanabe, Kurt Jaehnig, Sam Gabelt, John Bussan, Frank Lantz, George Winkert, and the WFF team. Micro-X operates under NASA Grant 80NSSC18K1445. Part of this work was performed under the auspices of the U.S. Department of Energy by Lawrence Livermore National Laboratory under Contract DE-AC52-07NA27344.

**Publisher's Note**  Springer Nature remains neutral with regard to jurisdictional claims in published maps and institutional affiliations.